\documentclass{article}

\usepackage{PRIMEarxiv}

\usepackage[utf8]{inputenc} 

\usepackage[T1]{fontenc}    
\usepackage{hyperref}       
\usepackage{url}            
\usepackage{booktabs}       
\usepackage{amsfonts}       
\usepackage{nicefrac}       
\usepackage{microtype}      
\usepackage{lipsum}
\usepackage{fancyhdr}       
\usepackage{graphicx}       
\usepackage{float}
\graphicspath{{media/}}     

\usepackage{scalerel,xparse}
\NewDocumentCommand\emoji{}{
    \scalerel*{
        \includegraphics{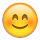}
    }{X}
}

\title{Predicting Gender and Political Affiliation Using Mobile Payment Data
}

\author{
  Ben Stobaugh \\
  Department of Computer Science \\
  The University of Texas at Austin \\
  \texttt{ben@helloben.co} \\
   \And
  Dhiraj Murthy \\
  Computational Media Lab \\
  Moody College of Communication \\
  The University of Texas at Austin \\
  \texttt{Dhiraj.Murthy@austin.utexas.edu} \\
}

\begin{document}
\maketitle

\begin{abstract}
We explore the understudied area of social payments to evaluate whether or not we can predict the gender and political affiliation of Venmo users based on the content of their Venmo transactions. Latent attribute detection has been successfully applied in the domain of studying social media. However, there remains a dearth of previous work using data other than Twitter. There is also a continued need for studies which explore mobile payments spaces like Venmo, which remain understudied due to the lack of data access. We hypothesize that using methods similar to latent attribute analysis with Twitter data, machine learning algorithms will be able to predict  gender and political affiliation of Venmo users with a moderate degree of accuracy. We collected crowdsourced training data that correlates participants' political views with their public Venmo transaction history through the paid Prolific service. Additionally, we collected 21 million public Venmo transactions from recently active users to use for gender classification. We then ran the collected data through a TF-IDF vectorizer and used that to train a support vector machine (SVM). After hyperparameter training and additional feature engineering, we were able to predict user's gender with a high level of accuracy (.91) and had modest success predicting user's political orientation (.63). 
\end{abstract}


\section{Introduction}
It has been previously shown that an individual’s purchasing history can predict various latent attributes about them \cite{zhang2020analysis}, such as their income, race, gender, and even political affiliation. Large retailers know this too and have been employing various methods to predict consumer attributes. For example, in 2012, Target used a woman’s purchasing history to send her unrequested baby-related advertising and a bevy of maternity-related coupons, revealing her pregnancy to her father \cite{kosinski2013private}. Similarly, Amazon uses purchase history to tailor product recommendations on a user-by-user basis \cite{zhou2022competing}. However, as researchers have shown, shopping habits can be used to predict more than just what we might buy next. Others were able to, with moderate accuracy, predict race, gender, income, and political affiliation from consumer data \cite{bertrand2018coming}. The implication of this is that our purchases can be mapped to demographic attributes and personal preferences. 

In the case of the United States, political culture during and post-Trump has become more polarized \cite{galvin2020party}. Moreover, such polarization can become part of consumer identity, associating particular brands, items, and stores with a political identity. For example, in 2012, Obama won 77\% of all counties that had a Whole Foods location, but only 29\% of counties that contained a Cracker Barrel \cite{wasserman}. In his book, “Why We’re Polarized”, Klein suggests the reason for this is that "[the companies] map onto our politics not because they are trying to serve one side of the political divide, but because our politics map onto our deeper preferences, and those deeper preferences drive much more than just our politics." \cite{klein2020we}. That is, in the case of the United States, there tend to be some general personality traits associated with being liberal and with being conservative. It is therefore possible to predict one’s political affiliation from short online posts, even if that affiliation is not explicit (e.g., attacking the leader of a political party).

Most latent attribute prediction related to social media uses data from Twitter, and to a lesser extent Facebook \cite{kosinski2013private, charlet2012detecting, rao2010classifying, pennacchiotti2011democrats, conover2011political, cohen2013classifying, pennacchiotti2011machine}. Most of this work relies on platform specific metrics such as hashtags or likes. Venmo, a US-based social payments platform, has several platform specific attributes which are not present in other social networks. These include factors such as whether a transaction was a “payment” or a “charge” or if the user we’re trying to predict was paid, known as the “target”, or created the payment, known as the “actor”. Moreover, users are required to include short text- and emoji- based ‘notes’ with their payments. These tend to consist of a couple of words, emoji, or a combination thereof. In this study, we seek to leverage Venmo data to identify whether trends seen on Twitter and Facebook are present in real purchases facilitated by the social spaces of the platform. Furthermore, we evaluate whether it is possible to use machine learned methods to predict the gender and political affiliation of Venmo users. Though evaluating the privacy implications and information leakage of Venmo posts is beyond the scope of this study, we frame this as an issue. 

Specifically, we extend and develop the work done by others using Twitter data to classify gender and political affiliation \cite{rao2010classifying, ardehaly2015inferring}, but with a unique type of social media data - Venmo posts. Venmo posts are qualitatively very different to both tweets and the latent attributes found in Twitter profile pages. Namely, Venmo posts are extremely brief and tend to employ a high percentage of emoji. We found that our gender model was extremely successful (.91), which, given there is a dearth of work on emoji-heavy text is a novel contribution to the literature. Our work also demonstrates gender classification accuracy better than Twitter-related studies (e.g., .83  \cite{culotta2016training}) and web forums (e.g., .86 \cite{zhang2011gender}).  Conversely, our political affiliation classifier was only moderately successful, with an accuracy of .63. This is less successful than most Twitter studies \cite{rao2010classifying, ardehaly2015inferring} but similar to Cohen’s and Ruth’s study \cite{cohen2013classifying} on “normal” users (.68) We partially attribute this to a lesser quantity of training data, but also from the lack of political identifiers found in Venmo transactions. Ultimately, through our crowd-sourced annotated data and machine learned models, our study confirms that the public data users are posting on Venmo can be used to successfully infer gender and, to more limited extent, political affiliation.

\section{Research Questions}
RQ1: Can gender be successfully predicted from sparse Venmo notes?

RQ2: Can political orientation be successfully predicted from sparse Venmo notes?

\section{Related Work}
\subsection{Latent Attribute Classification}
Though no study to date has explored political affiliation classification or gender prediction in the context of Venmo, studies have been conducted that explore these areas in the context of Facebook \cite{kosinski2013private} and, more extensively, Twitter \cite{rao2010classifying, pennacchiotti2011democrats, conover2011political, cohen2013classifying, pennacchiotti2011machine}. Additionally, some latent attribute analysis has been conducted on Venmo in the past, such as predicting user location \cite{yao2018beware} or identifying a user’s closest relationships \cite{zhang2017cold}.

Because of the high value of being able to identify an individual's political leanings, there is a rich literature on the topic of political affiliation classification. Indeed, researchers have been trying to use machine learning to solve this problem since before social media, with early studies looking at TV interview transcripts \cite{charlet2012detecting} and blog posts \cite{mullen2006preliminary}. These early studies proved the feasibility of using computational linguistics and machine learning to identify partisan political sentiment, as well as other traits such as gender and age from text. However, with the rise of both social media and off-the-shelf machine learning libraries lowering the barrier to entry, the literature has grown substantially over the past decade. One of the first to explore political orientation classification was Rao et. al.’s \cite{rao2010classifying} use of stacked-SVM-based classiﬁcation algorithms to determine the gender, age, and political leanings of a Twitter user. Soon after, other work emerged  that validated the methodology \cite{pennacchiotti2011democrats, conover2011political}. Namely, that it is possible to use tweet content to predict many different user attributes, including political affiliation. The accuracies in these papers range from 68\% \cite{cohen2013classifying} to 94\% \cite{mullen2006preliminary}. With the baseline established, researchers began to figure out where the limits in political affiliation classification lay. Volkova et. al. \cite{volkova2014inferring} examined the explicit tradeoff between accuracy and data quantity, finding that by using users’ neighbors they are able to achieve high accuracies with few users. Cohen and Ruths \cite{cohen2013classifying} argue that previous work tend to use data from “politically vocal” users and that published methods are nearly to 30\% less accurate when trained and tested on “normal” Twitter users. They selected their “normal” users by manually identifying Twitter users based on their tweet content, but avoiding politics enthusiasts. Giovanni et. al. \cite{di2018content} tested a variety of classifiers and tokenization methods and found that performing TF-IDF on nouns and then classifying with a Multilayer Perceptron yielded the highest results. 

\subsection{Venmo}

In April 2019, the US-based social payments platform Venmo had over 40 million users \cite{acker2020venmo}. By 2023, it had over 70 million users \cite{curry2021venmo}. Venmo exists in the intersection of payments and social media. Specifically, the platform allows users to publicly share transactions made between them and their peers \cite{acker2020venmo}. However, Venmo has come under criticism many times due to the fact that transactions are public unless deliberately changed in the settings. According to Venmo, this is because “it's fun to share [information] with friends in the social world" \cite{boyd_2018}. Critics, such as Mozilla and several researchers, have fired back, saying “by making privacy the default, Venmo can better protect […] users — and send a powerful message about the importance of privacy” \cite{verhage_2018}.

Early work such as Kraft et al.’s \cite{kraft2014security} "Security Research of a Social Payment App" documented several vulnerabilities in Venmo’s API, including one that allows anyone to see users’ "friends only" transactions. This was patched by Venmo before publication. The paper also highlights some social engineering tricks, most of which  have since been fixed on Venmo. Additionally, Kraft et. al \cite{kraft2014security} document Venmo’s unpublished, but public API, which we used during the course of this study to gather user transactions.

Other work on Venmo has highlighted the high level of personal information that can be derived when studying the history of an individual’s transactions on the platform. Khanna \cite{khanna2015venmo} created a Chrome Extension, Money Trail, which could be used to identify relationships between Venmo users and how much time they spend together. It can also identify members of private social organizations, attendees of private events, and users’ food purchases.

Similarly, Duc raised awareness of the public nature of Venmo transactions through her project “Public by Default” \cite{duc}, which uses transactions made public by the Venmo API to focus on 5 people and reveal highly personal information about their daily lives. For example, Duc was able to identify a couple living in Orange County, CA and find out where and how often they went to the movies or bought groceries. While Duc’s site shows just how much information can be learned about a person purely from their Venmo transactions, her findings are not scalable due to the manual effort that went into producing each one of her case studies. Regardless, the public pressure directed towards Venmo as a result of Duc’s project was swift and intense. The Mozilla foundation even started a campaign and a petition in order to get Venmo to change the default setting \cite{boyd_2018}. 

Yao et. al. \cite{yao2018beware} developed a series of algorithms (based loosely off of existing Twitter algorithms) to predict the home cities of Venmo users. The basic process can be broken down into four steps: 1) extract keywords from all of the transactions and divide them into four categories based on how reliable they are as an indicator of proximity; 2) construct a weighted undirected graph between users for each category where the weight is how often the two users transact; 3) identify “seed” users whose location is known and propagate throughout the graph; and 4) use the weights to combine the category graphs into one final graph with locations. By doing this, Yao et. al. \cite{yao2018beware} were able to predict the top-1, top-3, and top-5 possible locations for a Venmo user with accuracy up to 50\%, 80\%, and 90\%, respectively. 

In summary, there is a large amount of previous work on latent attribute prediction using major social media networks such as Facebook and Twitter, including exploring current limitations \cite{cohen2013classifying}. Though this body of work successfully classified gender and political affiliation on certain social media platforms, there is much room to extend these approaches to other domains. Because there is a particular dearth in the emergent area of social payments, which are ubiquitous in many countries, there is a critical need for latent attribute detection studies in this area. Venmo is an ideal platform to study as it has a public API and feed that resembles Twitter in certain ways, and there exists few latent detection studies specifically exploring Venmo data.

\section{Methodology}

We develop and extend previous successful methodologies deployed on Twitter and Facebook data \cite{rao2010classifying, di2018content}. One important item we had to consider while creating our study was how Venmo was both similar and dissimilar to Twitter and Facebook. The largest difference is that posts on Venmo are related to a payment, while posts on Facebook and Twitter can be about anything. This suggests that posts on Venmo are more representative of consumptive behaviors. Both Venmo and Twitter are known for their short and to-the-point messages. Unlike Twitter, Venmo’s limits are user-imposed rather than Twitter’s platform 280 character platform-imposed limit. Though Venmo posts can be 2,000 characters long, almost all Venmo transactions are very short. Indeed, we found that the most common post length was only one character (see figure 1). For our purposes, this means that the average Venmo post contains less features than the average Twitter or Facebook post.

\begin{figure}
\centering
\includegraphics[width=4in]{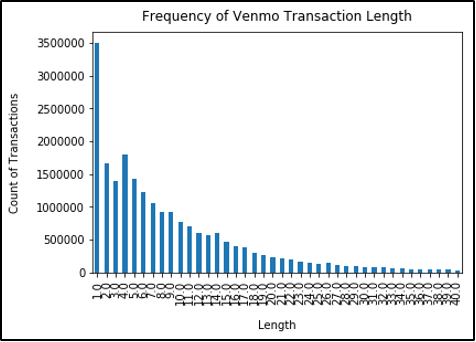}
\vspace{2pt}
\caption{Frequency of Venmo transaction character length in our dataset.}
\vspace{6pt}
\end{figure}

Furthermore, Venmo lacks many of the platform specific features of Twitter, such as mentions, hashtags, and retweets, but includes different ones we can take advantage of, which we discuss subsequently. Though Venmo has comments and likes, these are relatively minimally used, with only 2.7\% of transactions having comments and 11.3\% of transactions having likes \cite{zhang2017cold}. Our testing showed they did not meaningfully impact model performance. Our methodology consists of two main parts: data collection and data analysis.

\subsection{Data Collection}

Yao et. al. \cite{yao2018beware} were able to reliably predict user location by training on 500 of their 1,000 “seed” users. They gathered a ‘ground truth’ for location by associating Venmo users with their Facebook accounts. While some Venmo and Facebook accounts are linked, they also had to manually match profile pictures across the two services. They were then able to take the users’ Facebook location. For our purposes of gathering political leanings, this method is not ideal as we would be unable to account for the strength of participants’ views.

Instead, to collect political affiliation training data, we  employed a survey. This method allowed us to gather data in a format that is more suitable for machine learning tasks. As part of the survey, participants first give informed consent for us to use their data, then provide us with their Venmo username, basic demographic data such as age, education, sex, gender, and zip code. We then asked for information about their political leanings. We used a 10 question, 7 point measure of political affiliation, developed by Bail et. al. \cite{bail2018exposure}. 

We first tried using Amazon Mechanical Turk (MTurk). After running a pilot, we found that while we were able to gather a large amount of data in a relatively quick timeframe, the data received was almost entirely unreliable. Inconsistent answers on the political polling in addition to an unexpected amount of Venmo usernames being simply first names led us to this conclusion. This led us to try a different crowd-sourced service, Prolific. Because we wanted survey respondents to have at least 5 public Venmo transactions (a niche screening criteria), we set up our survey in two parts: one for screening and one for collecting demographics/training data. After taking the first survey, should the user provide a Venmo username and consent to having their transactions used, we use a script to gather all of their transactions. Venmo does not have an API endpoint to convert from usernames (provided in the survey) to user IDs (needed to gather all transactions). However, it does exist on a user’s profile page (https://venmo.com/username) as a JavaScript variable. To gather these in a programmatic way, we used Selenium to load the page, then executed JavaScript to return the user ID variable. Once we had the user ID, we used Venmo’s API to gather all of the user’s transactions. Due to the Venmo API requiring an access key from a valid Venmo account, the Computational Media Lab created a fictional account in order to access the API. We then gathered all of the Prolific user IDs associated with Venmo accounts that had greater than 5 public transactions. These users were then invited to take the second survey. After running a pilot, we found that approximately 20\% of users who completed the first survey qualified to take the second survey. Thus, in an attempt to receive 1,000 final responses, we attempted to solicit 5,000 responses to our qualification survey. We used Prolific’s prescreening options to request 2,500 Republican and 2,500 Democrat responses, in an attempt to minimize future class imbalances.

However, after 7 days we had only received 1,221 Republican responses, with only 218 of them having Venmo accounts with greater than 5 public transactions. This was less than half of our desired amount. We gathered a slightly larger amount of Democrat accounts, 346, but to avoid class imbalance only used 218 of them during training. We then solicited responses to our secondary survey in order to collect our training data. 

Additionally, we collected a large-scale dataset of Venmo transactions. To do this, we deployed a Venmo scraper in two phases. During the first phase, we simply collected transactions from Venmo’s “public feed” API. These are the posts that show up in the app when a user taps to see recent public posts. This is a slow process as the Venmo API endpoint only shows 20 transactions at a time and refreshes roughly every 15 minutes. However, through this method, we were able to collect 67,201 Venmo transactions, containing a total of 131,670 user IDs. Phase two was to collect every additional public transaction for each of these users. To do this, we developed a high throughput scraper for the Venmo API, taking advantage of  our University’s supercomputing cyberinfrastructure, in order to massively parallelize our work. This resulted in a dataset of 21,508,962 transactions.
 
We received IRB approval to collect and investigate this data. Each of our research team members has completed human subjects training and all data was stored securely using campus supercomputer infrastructure, which only authorized personnel could access.

\subsection{Data Analysis}

\begin{figure}[H]
\centering
\includegraphics[width=4in]{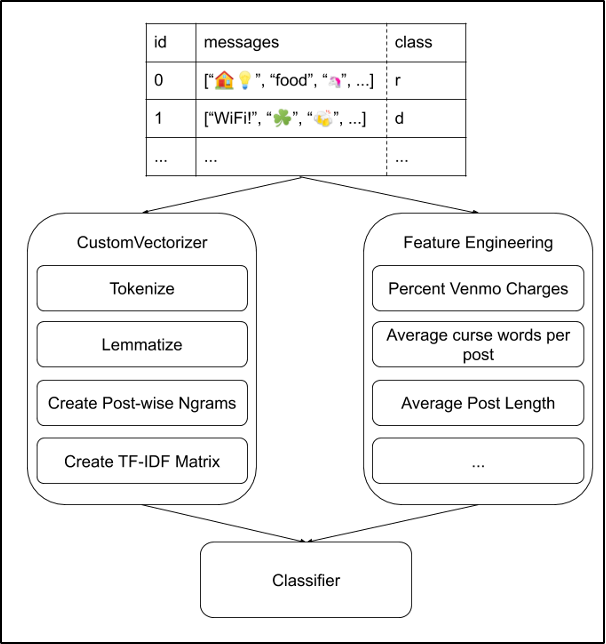}
\vspace{2pt}
\caption{Architecture of Venmo data analysis process.}
\vspace{6pt}
\end{figure} 

To analyze our data, we extended and developed existing latent attribute detection and political affiliation classification work done on Twitter (see figure 2). As such, we loosely followed the same procedures as Rao et. al. \cite{rao2010classifying} and Giovanni et. al. \cite{di2018content}. Although our models were tuned separately for gender and political affiliation, the process use for each was the same.

\subsubsection{Data Formatting}
The data scraper we developed provided us with a list of transactions, each containing an actor and target users. However, for data analysis we required a table of users, each containing associated transactions. Thus, our first step was to group the transactions by user ID and then extract important attributes such as transaction message and actor and/or target name. During this step we also aggregated attributes such as the percentage of transactions that are charges or the average number of likes each post has.

\subsubsection{Gender Ground Truth}
In order to determine the gender of our Venmo users to use as ground truth for prediction, we used a Python library called ‘gender-guesser’ \cite{arcos}. This library works by checking first names against a corpus of names in order to associate gender. The result will be one of ‘unknown’ (name not found), ‘andy’ (androgynous), ‘male’, ‘female’, ‘mostly\_male’, or ‘mostly\_female’. For our limited purposes, we drop any users that do not result in ‘male’ or ‘female’. Of course, there is no guarantee that user’s names on Venmo are reliable. To measure the accuracy of this method, we used our survey data where participants also provided their gender. Comparing the result of gender-guesser users’ Venmo names to the gender provided to us in the survey, we found that gender-guesser achieved slightly greater than 99\% accuracy when we specified the region as the United States. 

\subsubsection{Custom Vectorizer}
We built a custom vectorizer on top of Scikit-learn’s TfidfVectorizer and CountVectorizer \cite{pedregosa2011scikit} (we tried both). We decided to implement a custom vectorizer so that we could better control the tokenization/ngram generation steps compared to the built in models. We first created tokens from each post. SpaCy’s built-in tokenization best fit our needs for tokenizing both words and emoji. We then took the lemma of each word and used scikit-learn’s existing functionality to create ngrams. Through this approach, we were able to ensure that resulting ngrams are post-wise instead of just user-wise. That is, a word from one post will not end up in the ngram of a different post. This prevented transaction order from having an effect on our resulting vector space.

\subsubsection{Feature Engineering}
The features that we engineered fit into one of two separate categories. The first of these categories is structural features. These include: percentage of transactions that were initiated via charge (vs. payment), the average number of ‘likes’ received per post, their average post length, and the average length per post. The second category is content features. For these, we extended and developed from Rao. et. al. \cite{rao2010classifying}, including their list ‘socio-linguistic’ features. For each of these, we provided our model with the average number of feature occurrences per post as well as the percentage of a user’s posts which contain the feature. The features include: emoji (eg.\emoji), emoticons (eg. :-) ), Venmo emoji (eg. :uber:), words with repeated characters (eg. heyyyy), excitement (eg. !!!!), single exclaim (a single ! at the end of the post), ellipses (eg. …), shouting (the post is all uppercase), laughing (eg. lol, hahaha), oh my gods (eg. omg), and curse words.

\subsubsection{Classification}
We evaluated all classifiers that had been demonstrated to be successful in previous Twitter-based classification work. As such, we evaluated and tuned a Support Vector Machine (SVM), a simple Multilayer Perceptron (MLP), and a Gradient Boosted Decision Tree ensemble (GBDT). These evaluations were performed using a 5-fold cross validation and we tuned hyperparameters with a GridSearchCV for each. 

\section{Results}

After hyperparameter training and additional feature engineering, we were able to correctly identify gender with an accuracy of .91 and political orientation with an accuracy of .63. Though our political orientation model leaves room for improvement, this  results indicate success given that predicting this attribute from extremely sparse text is a far more challenging task than doing so with tweets.

We found that for both models, TFIDF outperformed a simple CountVectorizer, in contrast with some previous work \cite{rao2010classifying}. Additionally, our SVM had the highest accuracy of the classifiers we tried, justifying the majority of the classification media, but differing with some previous work \cite{di2018content}.

\begin{figure}
\centering
\includegraphics[width=6.5in]{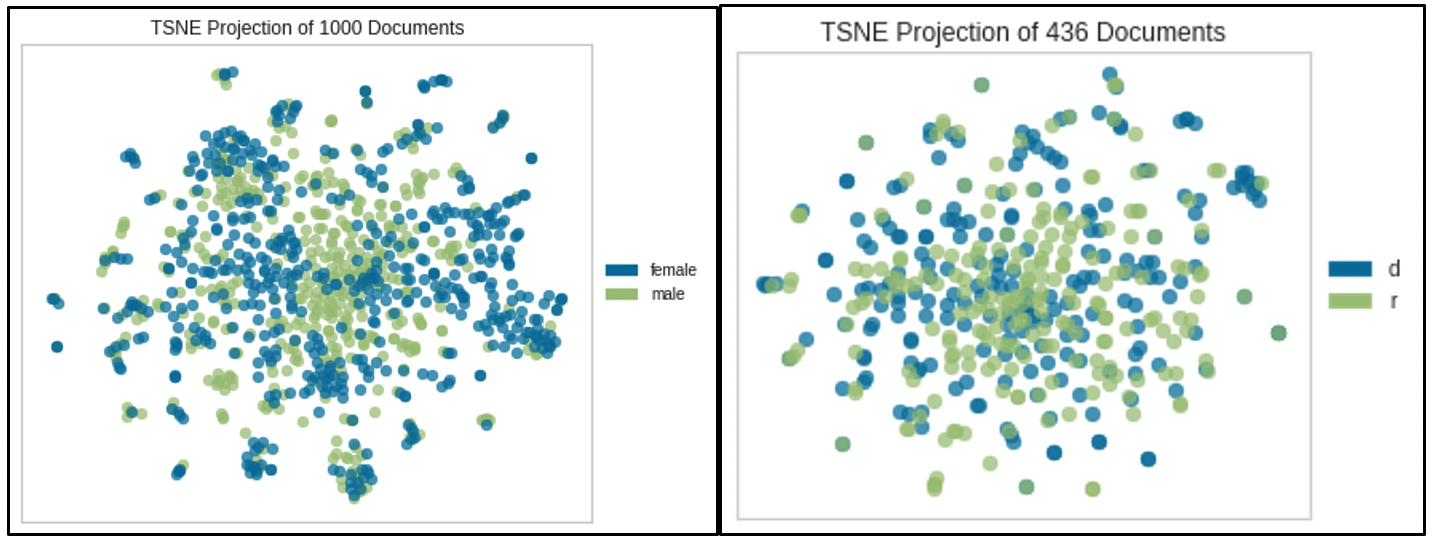}
\vspace{2pt}
\caption{ t-Distributed Stochastic Neighbor Embedding visualization plots of Venmo message n-grams generated for gender and political affiliation.}
\vspace{6pt}
\end{figure} 

\begin{figure}
\centering
\includegraphics[width=6.5in]{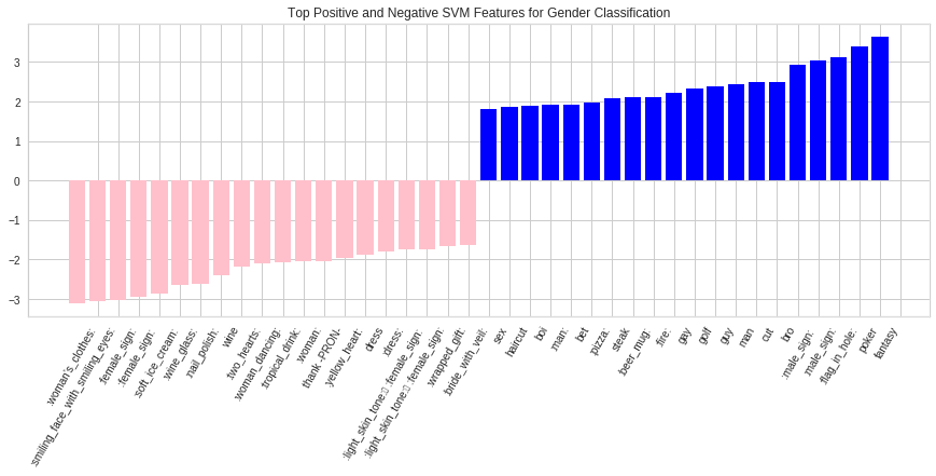}
\vspace{2pt}
\caption{The highest-weighted coefficients in our gender classification SVM \cite{arcos}; pink indicates coefficients for women and blue indicate coefficients for men; emoji are indicated with underscores}
\vspace{6pt}
\end{figure} 
As figures 3-4 indicate, the gender model was very successful. We attribute this to the fact that notes by men often had very gendered speech (e.g., Bro, man, boi). For women, we found that notes often reflected highly gendered activities, dress, and behaviors (e.g., dress emoji, high heels emoji, tropical drink emoji). It is interesting to note that gender for women had emoji as more important features than men (see figure 4). Both of these findings are in distinction to Twitter-based work, which  found lexical variety (e.g., gender markers are found with abbreviations like lol/omg, pronouns, and conjunctions).

\begin{figure}
\centering
\includegraphics[width=6.5in]{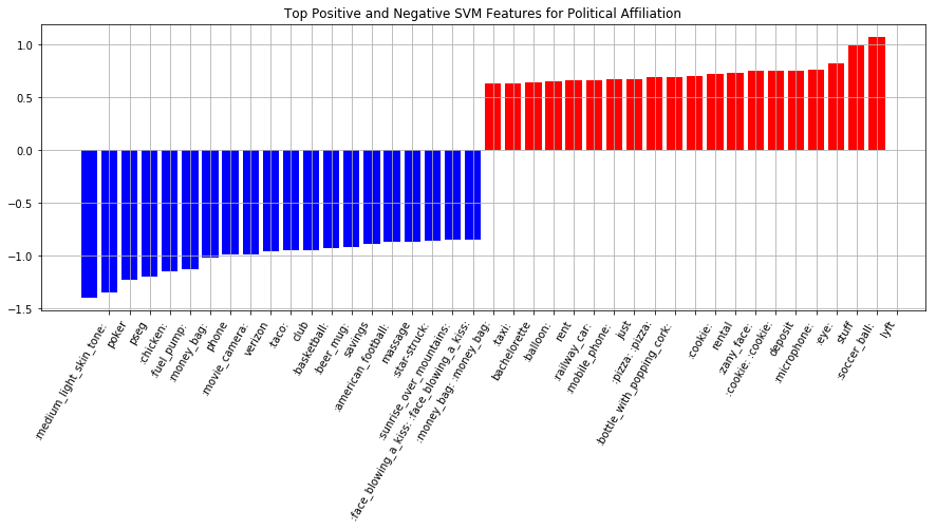}
\vspace{2pt}
\caption{The highest-weighted coefficients in our political affiliation classification SVM \cite{arcos}; red indicate coefficients for republican and blue indicate coefficients for democrat; emoji are indicated with underscores.}
\vspace{6pt}
\end{figure}

Figures 3-4 provide some indication of why the accuracy of our gender model was so high. Similarly, the lack of intuitive coefficients in figure 5 clearly indicates why our political affiliation model was not as successful. As mentioned previously, predicting political affiliation using social media data remains a challenging task as others have found \cite{cohen2013classifying}. Figure 5 provides a glimpse into the fact that Venmo posts failed to provide clear signals of Republican and Democrat affiliation. We believe that part of this noise in the data is due to the fact that Venmo users, particularly at the time of data collection, were heavily skewed towards the young adult demographic. In this sense, there was a higher level of convergence in terms of the types of consumption these users were engaged with. Moreover, the type of content being presented in their transaction notes is clearly not as indicative of political orientation as it is of gender. Moreover, predicting political orientation from such sparse text (e.g., a single emoji) is, unsurprisingly, a real challenge. Whereas for gender, single emoji such as high heels or lipstick had a much clearer signal for gender classification. However, for political orientation, a taco or a pizza are clearly not the best features for discerning political orientation. Moreover, we believe that detecting gender on Venmo is an easier task because, compared to Twitter, the percentage of users who use their own name is quite high. 

\section{Limitations and Future Work}
Although our results for political affiliation classification leave room for improvement, we do not believe that this proves such classification is not possible. We believe that the biggest obstacle holding our model back is data quantity. Volkova et. al. \cite{volkova2014inferring} showed that political affiliation classification is possible on Twitter with a small corpus of users if one takes advantage of the network of users around them; testing this on Venmo, however, is beyond the scope of our study. We believe that for better accuracy, the data could be expanded in two ways: 1) increasing the number of users in the dataset and 2) raising the minimum number of transactions per user. Additionally, we believe that future work could address the privacy/efficacy implications of Venmo maintaining public transactions by default in a landscape where third parties can infer so much. Future Venmo-based work should also explore diverse gender identities. We fully acknowledge that the gender classifier we used is limited to more traditional gender roles.  

\section{Conclusion}
In this study, we developed a method to accurately predict gender, and, to a lesser extent, political affiliation of Venmo users. Venmo is indeed a social media platform, which like Twitter, is home to large volumes of public social data, which can lead to the unintentional leakage of users’ latent attributes. We found that machine learning techniques for classification based purely on a user’s publicly posted transactions can be succesfully deployed on Venmo data. Particularly in light of the 2021 Facebook leaks \cite{larsen2022information}, what we disclose on social platforms remains critically important. Moreover, the efforts by technology companies in this area continues to be lacking. Other work using Venmo data makes clear that notes on Venmo can be used to discern demographic attributes or other sensitive information (e.g., participation in Alcoholics Anonymous groups \cite{tandon2022know}). This issue was made particularly clear when 
US President Joseph Biden’s Venmo account was found and his ‘friends’ seen \cite{tandon2022know}. Therefore, work such as ours which renders visible the extent of latent attribute identification is important to raising awareness.

\section*{Acknowledgments}
We thank Akaash Kolluri for providing manuscript feedback and assistance with formatting the manuscript and references for arXiv.

\bibliographystyle{unsrt}  
\bibliography{references}

\begin{thebibliography}{10}

\bibitem{zhang2020analysis}
Qian Zhang, Haruka Yamashita, Kenta Mikawa, and Masayuki Goto.
\newblock {Analysis of purchase history data based on a new latent class model
  for RFM analysis}.
\newblock {\em Industrial Engineering \& Management Systems}, 19(2):476--483,
  2020.

\bibitem{kosinski2013private}
Michal Kosinski, David Stillwell, and Thore Graepel.
\newblock {Private traits and attributes are predictable from digital records
  of human behavior}.
\newblock {\em Proceedings of the national academy of sciences},
  110(15):5802--5805, 2013.

\bibitem{zhou2022competing}
Bo~Zhou and Tianxin Zou.
\newblock {Competing for recommendations: The strategic impact of personalized
  product recommendations in online marketplaces}.
\newblock {\em Marketing Science}, 2022.

\bibitem{bertrand2018coming}
Marianne Bertrand and Emir Kamenica.
\newblock {Coming apart? Cultural distances in the United States over time}.
\newblock Technical report, National Bureau of Economic Research, 2018.

\bibitem{galvin2020party}
Daniel~J Galvin.
\newblock {Party domination and base mobilization: Donald Trump and Republican
  Party building in a polarized era}.
\newblock In {\em The Forum}, volume~18, pages 135--168. De Gruyter, 2020.

\bibitem{wasserman}
D.~Wasserman.
\newblock {The Cook Political Report’s 2014 Election Road Map}.
\newblock 2014.
\newblock Arlington, VA.

\bibitem{klein2020we}
Ezra Klein.
\newblock {\em {Why we're polarized}}.
\newblock Simon and Schuster, 2020.

\bibitem{charlet2012detecting}
Delphine Charlet and G{\'e}raldine Damnati.
\newblock {Detecting politician speech in TV broadcast news shows}.
\newblock In {\em 2012 10th International Workshop on Content-Based Multimedia
  Indexing (CBMI)}, pages 1--6. IEEE, 2012.

\bibitem{rao2010classifying}
Delip Rao, David Yarowsky, Abhishek Shreevats, and Manaswi Gupta.
\newblock {Classifying latent user attributes in twitter}.
\newblock In {\em Proceedings of the 2nd international workshop on Search and
  mining user-generated contents}, pages 37--44, 2010.

\bibitem{pennacchiotti2011democrats}
Marco Pennacchiotti and Ana-Maria Popescu.
\newblock {Democrats, republicans and starbucks afficionados: user
  classification in twitter}.
\newblock In {\em Proceedings of the 17th ACM SIGKDD international conference
  on Knowledge discovery and data mining}, pages 430--438, 2011.

\bibitem{conover2011political}
Michael Conover, Jacob Ratkiewicz, Matthew Francisco, Bruno Gon{\c{c}}alves,
  Filippo Menczer, and Alessandro Flammini.
\newblock {Political polarization on twitter}.
\newblock In {\em Proceedings of the international aaai conference on web and
  social media}, volume~5, pages 89--96, 2011.

\bibitem{cohen2013classifying}
Raviv Cohen and Derek Ruths.
\newblock {Classifying political orientation on Twitter: It’s not easy!}
\newblock In {\em Proceedings of the International AAAI Conference on Web and
  Social Media}, volume~7, pages 91--99, 2013.

\bibitem{pennacchiotti2011machine}
Marco Pennacchiotti and Ana-Maria Popescu.
\newblock {A machine learning approach to twitter user classification}.
\newblock In {\em Proceedings of the international AAAI conference on web and
  social media}, volume~5, pages 281--288, 2011.

\bibitem{ardehaly2015inferring}
Ehsan~Mohammady Ardehaly and Aron Culotta.
\newblock {Inferring latent attributes of Twitter users with label
  regularization}.
\newblock In {\em Proceedings of the 2015 conference of the north american
  chapter of the association for computational linguistics: Human language
  technologies}, pages 185--195, 2015.

\bibitem{culotta2016training}
Aron Culotta.
\newblock {Training a text classifier with a single word using Twitter Lists
  and domain adaptation}.
\newblock {\em Social Network Analysis and Mining}, 6:1--15, 2016.

\bibitem{zhang2011gender}
Yulei Zhang, Yan Dang, and Hsinchun Chen.
\newblock {Gender classification for web forums}.
\newblock {\em IEEE Transactions on Systems, Man, and Cybernetics-Part A:
  Systems and Humans}, 41(4):668--677, 2011.

\bibitem{yao2018beware}
Xin Yao, Yimin Chen, Rui Zhang, Yanchao Zhang, and Yaping Lin.
\newblock {Beware of what you share: Inferring user locations in Venmo}.
\newblock {\em IEEE Internet of Things Journal}, 5(6):5109--5118, 2018.

\bibitem{zhang2017cold}
Xinyi Zhang, Shiliang Tang, Yun Zhao, Gang Wang, Haitao Zheng, and Ben Zhao.
\newblock {Cold hard e-cash: Friends and vendors in the venmo digital payments
  system}.
\newblock In {\em Proceedings of the International AAAI Conference on Web and
  Social Media}, volume~11, pages 387--396, 2017.

\bibitem{mullen2006preliminary}
Tony Mullen and Robert Malouf.
\newblock {A Preliminary Investigation into Sentiment Analysis of Informal
  Political Discourse.}
\newblock In {\em AAAI spring symposium: computational approaches to analyzing
  weblogs}, pages 159--162, 2006.

\bibitem{volkova2014inferring}
Svitlana Volkova, Glen Coppersmith, and Benjamin Van~Durme.
\newblock {Inferring user political preferences from streaming communications}.
\newblock In {\em Proceedings of the 52nd Annual Meeting of the Association for
  Computational Linguistics (Volume 1: Long Papers)}, pages 186--196, 2014.

\bibitem{di2018content}
Marco Di~Giovanni, Marco Brambilla, Stefano Ceri, Florian Daniel, and Giorgia
  Ramponi.
\newblock {Content-based classification of political inclinations of Twitter
  users}.
\newblock In {\em 2018 IEEE International Conference on Big Data (Big Data)},
  pages 4321--4327. IEEE, 2018.

\bibitem{acker2020venmo}
Amelia Acker and Dhiraj Murthy.
\newblock {What is Venmo? A descriptive analysis of social features in the
  mobile payment platform}.
\newblock {\em Telematics and Informatics}, 52:101429, 2020.

\bibitem{curry2021venmo}
David Curry.
\newblock {Venmo revenue and usage statistics}.
\newblock {\em Business of Apps}, 2023.

\bibitem{boyd_2018}
Ashley Boyd.
\newblock {25,000 Americans Urge Venmo to update its privacy settings}.
\newblock {\em The Mozilla Blog}, 2018.

\bibitem{verhage_2018}
Julie Verhage.
\newblock {Venmo considers making it harder to see what other people are
  buying}.
\newblock {\em Bloomberg News}, Aug 2018.

\bibitem{kraft2014security}
Ben Kraft, Eric Mannes, and Jordan Moldow.
\newblock {Security research of a social payment app}.
\newblock {\em Citeseer}, 2014.

\bibitem{khanna2015venmo}
Aran Khanna.
\newblock {Venmo’ed: Sharing your payment data with the world}.
\newblock {\em Technology Science}, 2015102901, 2015.

\bibitem{duc}
Hang Do~Thi Duc.
\newblock {Public by default}, 2018.
\newblock Available from: https://publicbydefault.fyi/.

\bibitem{bail2018exposure}
Christopher~A Bail, Lisa~P Argyle, Taylor~W Brown, John~P Bumpus, Haohan Chen,
  MB~Fallin Hunzaker, Jaemin Lee, Marcus Mann, Friedolin Merhout, and Alexander
  Volfovsky.
\newblock {Exposure to opposing views on social media can increase political
  polarization}.
\newblock {\em Proceedings of the National Academy of Sciences},
  115(37):9216--9221, 2018.

\bibitem{arcos}
D.~Arcos.
\newblock {Lead-Ratings/Gender Guesser}.
\newblock 2016.

\bibitem{pedregosa2011scikit}
Fabian Pedregosa, Ga{\"e}l Varoquaux, Alexandre Gramfort, Vincent Michel,
  Bertrand Thirion, Olivier Grisel, Mathieu Blondel, Peter Prettenhofer, Ron
  Weiss, Vincent Dubourg, et~al.
\newblock {Scikit-learn: Machine learning in Python}.
\newblock {\em the Journal of machine Learning research}, 12:2825--2830, 2011.

\bibitem{larsen2022information}
Rebekah Larsen.
\newblock {‘Information Pressures’ and the Facebook Files: Navigating
  Questions around Leaked Platform Data}.
\newblock {\em Digital Journalism}, 10(9):1591--1603, 2022.

\bibitem{tandon2022know}
Rajat Tandon, Pithayuth Charnsethikul, Ishank Arora, Dhiraj Murthy, and Jelena
  Mirkovic.
\newblock {I know what you did on Venmo: Discovering privacy leaks in mobile
  social payments}.
\newblock {\em Proceedings on Privacy Enhancing Technologies}, 3:200--221,
  2022.

\end{thebibliography}

\end{document}